\documentclass[12pt]{article}
\usepackage[top=30mm, bottom=40mm, left=25mm, right=25mm]{geometry}
\usepackage{cite} 

\usepackage{graphicx}
\usepackage{amsmath, amssymb} 				
\usepackage{empheq}
\usepackage{fancybox}
\usepackage{mathrsfs} 
\usepackage{textcomp} 
\usepackage{mathtools}

\usepackage{epstopdf}

\usepackage{latexsym}
\usepackage{marvosym}
\usepackage{bm}	

\usepackage[mathcal]{euscript}
\usepackage{indentfirst}

\usepackage{hyperref}
\hypersetup{
	unicode,	
	colorlinks,
	citecolor=cyan,
	linkcolor=cyan, 
	bookmarksopen=true,
	bookmarksopenlevel=\maxdimen,
	bookmarksnumbered
	}

\numberwithin{equation}{section}

\def\K3{\mathrm K3}

\def\gl#1#2{$\mathrm{GL}(#1; {\bf #2})$}
\def\sl#1#2{$\mathrm{SL}(#1; {\bf #2})$}
\def\sp#1#2{$\mathrm{Sp}(#1; {\bf #2})$}
\def\spin#1#2{$\mathrm{Spin}(#1, #2)$}
\def\su#1#2{SU({#1,#2})}
\def\usp#1#2{USp({#1,#2})}
\def\U#1{U({#1})}

\def\O#1{O({#1})}

\def\double #1{#1{\hbox{\kern-2pt $#1$}}}
\def\un#1{\underline #1}

\def\gl#1#2{\ifmmode \mathrm{GL}(#1; {\bf #2}) \else $\mathrm{GL}(#1; {\bf #2})$\fi}
\def\sl#1#2{\ifmmode \mathrm{SL}(#1; {\bf #2}) \else $\mathrm{SL}(#1; {\bf #2})$\fi}
\def\so#1{\ifmmode \mathrm{SO}({#1}) \else $\mathrm{SO}(#1)$\fi}

\def\sp#1#2{\ifmmode \mathrm{Sp}(#1; {\bf #2}) \else $\mathrm{Sp}(#1; {\bf #2})$\fi}
\def\usp#1#2{\ifmmode \mathrm{USp}(#1,#2) \else $\mathrm{USp}(#1,#2)$\fi}
\def\spin#1#2{\ifmmode \mathrm{Spin}(#1,#2) \else $\mathrm{Spin}(#1,#2)$\fi} 
\def\su#1{\ifmmode \mathrm{SU}({#1}) \else $\mathrm{SU}(#1)$\fi}

\def\Gamma{\mathchar"0100}
\def\Delta{\mathchar"0101}
\def\Theta{\mathchar"0102}
\def\Lambda{\mathchar"0103}
\def\Xi{\mathchar"0104}
\def\Pi{\mathchar"0105}
\def\Sigma{\mathchar"0106}
\def\Upsilon{\mathchar"0107}
\def\Phi{\mathchar"0108}
\def\Psi{\mathchar"0109}
\def\Omega{\mathchar"010A}

\def\on#1#2{{\buildrel{\mkern2.5mu#1\mkern-2.5mu}\over{#2}}}
\def\dt#1{\on{\hbox{\bf .}}{#1}}                
\def\under#1#2{\mathop{\null#2}\limits_{#1}}	
\mathcode`\*="702A                  
\def\f#1#2{{\textstyle{#1\over#2}}}	   
\def\half{{\textstyle{1\over{\raise.1ex\hbox{$\scriptstyle{2}$}}}}}
\def\slap#1#2{\setbox0=\hbox{$#1{#2}$}#2\kern-\wd0{\hbox to\wd0{\hfil$#1{/}$\hfil}}}
\def\sla#1{\mathpalette\slap{#1}}		

\catcode128=13 \def €{\"A}                 
\catcode129=13 \def {\AA}                 
\catcode130=13 \def '{\c}           	   
\catcode131=13 \def ƒ{\'E}                   
\catcode132=13 \def "{\~N}                   
\catcode133=13 \def …{\"O}                 
\catcode134=13 \def †{\"U}                  
\catcode135=13 \def ‡{\'a}                  
\catcode136=13 \def ˆ{\`a}                   
\catcode137=13 \def ‰{\^a}                 
\catcode138=13 \def Š{\"a}                 
\catcode139=13 \def ‹{\~a}                   
\catcode140=13 \def Œ{\alpha}            
\catcode141=13 \def {\chi}                
\catcode142=13 \def Ž{\'e}                   
\catcode143=13 \def {\`e}                    
\catcode144=13 \def {\^e}                  
\catcode145=13 \def '{\"e}                
\catcode146=13 \def '{\'\i}                 
\catcode147=13 \def "{\`\i}                  
\catcode148=13 \def "{\^\i}                
\catcode149=13 \def •{\"\i}                
\catcode150=13 \def –{\~n}                  
\catcode151=13 \def —{\'o}                 
\catcode152=13 \def ˜{\`o}                  
\catcode153=13 \def ™{\^o}                
\catcode154=13 \def š{\"o}                 
\catcode155=13 \def ›{\~o}                  
\catcode156=13 \def œ{\'u}                  
\catcode157=13 \def {\`u}                  
\catcode158=13 \def ž{\^u}                
\catcode159=13 \def Ÿ{\"u}                
\catcode160=13 \def  {\tau}               
\catcode162=13 \def ¢{\oplus}           
\catcode163=13 \def £{\relax\ifmmode\expandafter\to\else\expandafter\itemize\fi} 
\catcode164=13 \def ¤{\subset}	  
\catcode165=13 \def ¥{\infty}           
\catcode166=13 \def ¦{\mp}                
\catcode167=13 \def §{\sigma}           
\catcode168=13 \def ¨{\gamma}               
\catcode169=13 \def ©{\gamma}         
\catcode170=13 \def ª{\leftrightarrow} 
\catcode171=13 \def «{\relax\ifmmode\expandafter\acute\else\expandafter\'\fi}
\catcode172=13 \def ¬{\relax\ifmmode\expandafter\ddt\else\expandafter\"\fi}
\catcode173=13 \def ­{\equiv}            
\catcode174=13 \def ®{{{}={}}}          
\catcode175=13 \def ¯{\Omega}          
\catcode176=13 \def °{\otimes}          
\catcode177=13 \def ±{\ne}                 
\catcode178=13 \def ²{\le}                   
\catcode179=13 \def ³{\ge}                  
\catcode180=13 \def ´{\upsilon}          
\catcode181=13 \def µ{\alpha}                
\catcode182=13 \def ¶{\delta}             
\catcode183=13 \def ·{\epsilon}          
\catcode184=13 \def ¸{\Pi}                  
\catcode185=13 \def ¹{\pi}                  
\catcode186=13 \def º{\beta}               
\catcode187=13 \def »{\partial}           
\catcode188=13 \def ¼{\nobreak\ }       
\catcode189=13 \def ½{\zeta}               
\catcode190=13 \def ¾{\sim}                 
\catcode191=13 \def ¿{\omega}           
\catcode192=13 \def À{\dt}                     
\catcode193=13 \def Á{\gets}                
\catcode194=13 \def Â{\lambda}           
\catcode195=13 \def Ã{\beta}                   
\catcode196=13 \def Ä{\phi}                  
\catcode197=13 \def Å{\xi}                     
\catcode198=13 \def Æ{\psi}                  
\catcode199=13 \def Ç{\int}                    
\catcode200=13 \def È{\oint}                 
\catcode201=13 \def É{\relax\ifmmode\expandafter\cdot\else\vol\fi}    
\catcode202=13 \def Ê{\relax\ifmmode\expandafter\,\else\thinspace\fi}
\catcode203=13 \def Ë{\`A}                      
\catcode204=13 \def Ì{\~A}                      
\catcode205=13 \def Í{\~O}                      
\catcode206=13 \def Î{\Theta}              
\catcode207=13 \def Ï{\theta}               
\catcode208=13 \def Ð{\relax\ifmmode\expandafter\bar\else\expandafter\=\fi}
\catcode209=13 \def Ñ{\overline}             
\catcode210=13 \def Ò{\langle}               
\catcode211=13 \def Ó{\relax\ifmmode\expandafter\{\else\ital\fi}      
\catcode212=13 \def Ô{\rangle}               
\catcode213=13 \def Õ{\}}                        
\catcode214=13 \def Ö{\sla}                      
\catcode215=13 \def ×{\relax\ifmmode\expandafter\check\else\expandafter\v\fi}
\catcode216=13 \def Ø{\"y}                     
\catcode217=13 \def Ù{\"Y}  		    
\catcode218=13 \def Ú{\Leftarrow}       
\catcode219=13 \def Û{\Leftrightarrow}       
\catcode220=13 \def Ü{\relax\ifmmode\expandafter\Rightarrow\else\sect\fi}
\catcode221=13 \def Ý{\sum}                  
\catcode222=13 \def Þ{\prod}                 
\catcode223=13 \def ß{\widehat}              
\catcode224=13 \def à{\pm}                     
\catcode225=13 \def á{\nabla}                
\catcode226=13 \def â{\quad}                 
\catcode227=13 \def ã{\in}               	
\catcode228=13 \def ä{\star}      	      
\catcode229=13 \def å{\sqrt}                   
\catcode230=13 \def æ{\^E}			
\catcode231=13 \def ç{\Upsilon}              
\catcode232=13 \def è{\"E}    	   	 
\catcode233=13 \def é{\`E}               	  
\catcode234=13 \def ê{\Sigma}                
\catcode235=13 \def ë{\Delta}                 
\catcode236=13 \def ì{\Phi}                     
\catcode237=13 \def í{\`I}        		   
\catcode238=13 \def î{\iota}        	     
\catcode239=13 \def ï{\Psi}                     
\catcode240=13 \def ð{\times}                  
\catcode241=13 \def ñ{\Lambda}             
\catcode242=13 \def ò{\cdots}                
\catcode243=13 \def ó{\^U}			
\catcode244=13 \def ô{\`U}    	              
\catcode245=13 \def õ{\bo}                       
\catcode246=13 \def ö{\relax\ifmmode\expandafter\hat\else\expandafter\^\fi}
\catcode247=13 \def÷{\relax\ifmmode\expandafter\tilde\else\expandafter\~\fi}
\catcode248=13 \def ø{\ll}                         
\catcode249=13 \def ù{\gg}                       
\catcode250=13 \def ú{\eta}                      
\catcode251=13 \def û{\kappa}                  
\catcode252=13 \def ü{\half}     		 
\catcode253=13 \def ý{\Gamma} 		
\catcode254=13 \def þ{\Xi}   			
\catcode255=13 \def ÿ{\relax\ifmmode\expandafter{}^{\dagger}{}\else\dag\fi}

\def\rgboo#1{\pdfliteral{#1 rg #1 RG}}
\def\rgb#1#2{\rgboo{#1}#2\rgboo{0 0 0}}

\def\A{{\cal A}}  \def\B{{\cal B}}  \def\C{{\cal C}}  \def\D{{\cal D}}
\def\E{{\cal E}}         
  \def\K{{\cal K}}         
\def\O{{\cal O}}  \def\P{{\cal P}}    
\def\S{{\cal S}}  \def\T{{\cal T}}  \def\U{{\cal U}} \def\V{{\cal V}}  \def\W{{\cal W}}

\def\^{\wedge}

\def\dd{\hbox{\,\Large$\triangleright$}}

\def\dig#1{\setbox0=\hbox{$#1M$}
	\hskip.06\wd0 \vrule width.07\wd0 height.63\wd0 depth.01\wd0 
	\vrule width.37\wd0 height.63\wd0 depth-.56\wd0 \hskip-.4\wd0
	\vrule width.25\wd0 height.35\wd0 depth-.28\wd0 
	\vrule width.07\wd0 height.35\wd0 depth-.17\wd0 \hskip.14\wd0}
\def\digamma{{\mathpalette\dig{}}}
\def\di{\digamma}

\title{
\Huge\bfseries\sffamily \strut\rgb{1 0 .3}{Critical Super F-theories}
}
\author{William D. Linch \textsc{iii}$\,{}^\text{\Pisces}$ and Warren Siegel$\,{}^\text{\Scorpio}$}
\date{}		

\begin{document}
\maketitle

\vspace*{-65mm}
\begin{flushright}
{
UMDEPP-015-010\\
YITP-SB-15-23
}
\end{flushright}
\vspace*{+40mm}

\begin{center}
{\em
${}^{\mbox{\footnotesize\Pisces}}$
Center for String and Particle Theory,
Department of Physics,\\
University of Maryland at College Park,
College Park, MD 20742-4111\\
~\\
${}^{\mbox{\footnotesize\Scorpio}}$
C. N. Yang Institute for Theoretical Physics\\
State University of New York, Stony Brook, NY 11794-3840
}
\end{center} 

\vspace{10pt}

\begin{abstract}
We present F-theories that reduce to 10D Type II Green-Schwarz superstrings.  They vary in manifest U-duality according to division between spacetime and ``internal" coordinates.  They are defined by selfdual current superalgebras in higher worldvolume dimensions with manifest $\mathrm G\times \mathrm G'$ symmetry where the spacetime symmetry $\mathrm G=\mathrm E_{n(n)}$ ranges over the (split form of the) exceptional groups with ranks $n=\mathrm D+1 \leq 7$ and the internal symmetry $\mathrm {G'= GL(10-D)}$. 
\end{abstract}

\vspace*{.5cm}
\begin{flushleft}
~\\
{${}^{\mbox{\footnotesize\Pisces}}$ \href{mailto:wdlinch3@gmail.com}{wdlinch3@gmail.com}}\\
{$^{\text{\Scorpio}}$ \href{mailto:siegel@insti.physics.sunysb.edu}{siegel@insti.physics.sunysb.edu}}
\end{flushleft}

\setcounter{page}0
\thispagestyle{empty}
\newpage

{\linespread{0.8}\tableofcontents}

\section{Prologue}

Attempts to describe fundamental ``membranes" (and higher-dimensional worldvolumes) date back to the work of Dirac \cite{Dirac:1962iy}. Such models
apparently suffer from spectral difficulties, such as continuous mass or lack of massless states.  
A major problem there is the nature of the worldvolume metric.  A useful way to view this might be to compare to strings and particles.  For this purpose we consider Polyakov's action \cite{Polyakov:1986cs}
\begin{equation}
L_P = Â^{mn}[(»_m X)É(»_n X) - g_{mn}] + \f1{Œ'}å{-g} ,
\end{equation}
generalized from the string. Variation with respect to $Â$ identifies $g_{mn}$ with the induced metric, yielding an action proportional to the induced worldvolume:  
\begin{equation}
L_{geo} = \f1{Œ'}å{-\mathrm{det}[(»_m X)É(»_n X)]}  .
\end{equation}
This is the Nambu-Goto action for the string, but for the particle it describes the case with mass $1/Œ'$.  For the membrane, this is the form of the action that is generally assumed by analogy to the string; but perhaps the analogy to the particle is more appropriate.  Supersymmetry won't necessarily help:  Massive superparticles can be derived by dimensional reduction with central charges; this reduction produces a cosmological term.  They have $û$ symmetry of the form $¶Ï=(1+Œ'Öp)û$, similar to that of the supermembrane \cite{Bergshoeff:1987cm} which may also lack massless states \cite{deWit:1988ig, Mezincescu:1987kj}.

The string is exceptional for its Weyl scale invariance.  This is seen from the other derived form of the action:  Varying $L_P$ with respect to $g_{mn}$ gives
\begin{equation}
Â^{mn} = \f1{Œ'}üå{-g}g^{mn} .
\end{equation}
Substitution back into the action (at least classically) yields
\begin{equation}
L_{BdVHDZ} = \f1{Œ'}ü\left[ å{-g}g^{mn}(»_m X)É(»_n X) -({\rm d}-2)å{-g}Ê\right]
\end{equation}
for d worldvolume dimensions.  This is the Brink-diÊVecchia-Howe-Deser-Zumino action \cite{Brink:1976sc,Deser:1976rb} for the string, plus a cosmological term for all other cases \cite{Howe:1977hp}.  (However, Weyl invariant actions have been considered that are not quadratic in $X$ even with an independent worldvolume metric \cite{Lindstrom:1987cv}.) 
The effect of a cosmological term for the closed bosonic string (Liouville theory) is to produce a ``continuous spectrum" (cut) \cite{Curtright:1982gt} (as for subcritical closed strings without the term \cite{Lovelace:1971fa}).

In references \cite{Linch:2015fya,Linch:2015qva} we initiated a study of fundamental branes, the quantization of which is proposed to define lower-dimensional F-theories beyond their supergravity approximation. 
Since our F-theories should reduce to standard Green-Schwarz superstrings upon solving dimensional reduction and section conditions, we expect our formalism to avoid the problems described above.  
At this point we know several advantages over conventional approaches to fundamental superbranes: (1) Reduction to the string is imposed by the formalism, not found as a limit.  
Thus it contains the same degrees of freedom as the usual strings, only more 0-modes, such as winding modes, are evident.
(2) More U duality can be manifested.  (3) Perhaps most importantly, it has worldvolume conformal invariance (at least for the selfdual differential forms $X$) which would seem to preclude the appearance of worldvolume cosmological terms.

In the next section we give an introduction to the method by considering the F-theories of bosonic strings in D=1 and 2 spacetime dimensions.  All our F-theories are based on the principle that all the bosonic worldvolume fields should be selfdual (in the sense of T-duality), perhaps by including their duals.
This is followed by a qualitative definition of the general method and its purpose, including supersymmetry and the completion of all F-theories to 10D superstrings by including ``internal" dimensions (scalar worldvolume fields) $Y$ and their ``duals" $\widetilde Y$.
After a brief section on notation, we give the general type of current superalgebras on which the Hamiltonian version of the formalism is based. 
These superalgebras are a generalization of one handedness of the Green-Schwarz superstring to selfdual supercurrents in higher worldvolume dimensions.
The bosons are then separated into $X,Y,\widetilde Y$.
We give a derivation of constraints in section \ref{S:Constraints}.
We then get into the details of the different dimensions, specified by the particular values of the structure constants, which define supersymmetry, and of the generalized metric found in the bosonic theory, needed for the Fierz-type identities that fix the total spacetime dimension classically to 3,4,6,10.
We conclude with our Prospects for future work.

\section{Lowest Dimensions}
\label{limbo}

As a ``review" of the bosonic formalism, 
in the following subsections (labeled \ref{limbo}.D) 
we begin by presenting the simple examples of the worldvolume gauge theories generating the F-theories corresponding to the bosonic string in D=1 and 2 spacetime dimensions; the cases D=3 and 4 were treated in references \cite{Linch:2015fya,Linch:2015qva}. 
All 4 cases turn out to be chiral gauge forms $X$ on the d-dimensional worldvolume \cite{Siegel:1983es} (or their odd-d analogue in the case of D=1).

The fieldstrengths $F=dX$ are split into (anti-)selfdual parts $F^{(\pm)} = (1\pm \ast) F$ in terms of which the Lagrangian actions are presented in a ``conformal gauge'' 
\begin{align}
\label{E:LagrangianAction}
S =  \tfrac12 \int F^{(+)} \cdot F^{(-)} \, d^{Ê\rm d} \sigma 
\end{align}
in which the worldvolume metric (``\,$\cdot$\,'' denotes contraction using this metric) and Lagrange multiplier for self-duality have been gauge-fixed. The symmetry group of the Lagrangian description is denoted by L.

The momenta ${P}$ conjugate to $X$ are defined by ${P} := - \delta S/\delta \dt X$. Due to the form of the action, they are always the $\tau$ component of the fieldstrength $F$. 
Underlined lower-case Roman indices in this section will refer to worldvolume indices (e.g., $\sigma^{\un a}$) in their Lagrangian form. In their Hamiltonian form, this is reduced $\sigma^{\un a}\to \tau , \sigma^a$ to a time-like parameter $\tau$ and the remaining d$-$1 worldvolume parameters $\sigma^a$, reducing the symmetry L $\to$ H. The action is given in Hamiltonian form by 
\begin{align}
\label{E:HamiltonianAction}
S = -\int {P}\cdot \dt X \, d^{Ê\rm d-1} \sigma d\tau + \int H\, d\tau
\end{align}
in terms of the Hamiltonian $H$.
Analogously to electromagnetism, the $\tau$ component of the gauge field $X$ becomes the Lagrange multiplier for the Gau\ss{} law constraint $\U \sim \partial {P}$.

In each case, the invariant currents $\P$ ($\tilde \P$) are the $\tau$ components of the (anti-)self-dual parts of the gauge fieldstrength. Due to the uniformity of the theories, these are all of the form $\P = {P} + \eta_a \partial^a X$ and $\tilde \P = {P} - \eta_a \partial^a X$ for some invariant tensor $ú$.
Consequently, their quantum brackets are all of the form $[\P(1), \P(2)] = 2i\eta_a \partial^a \delta$.
The symmetry group of these currents and their bracket algebra is strictly larger than H; we will call it G.
The $P$'s and the quantum brackets they imply are given explicitly in each case. 
(The groups L, H, and G are summarized (and extended to $\mathrm D > 2$) in table \ref{T:structure}.)

\subsection{1D String}
\label{limbo.1}
This system results from the Hamiltonian analysis of a worldvolume gauge 1-form $X_{\un a}$ and a 0-form $X$ on a worldvolume modeled on $\mathbf R^{2,1}$ with 
$\mathrm L = \sl2R$-covariant
fieldstrengths
\begin{align}
F^{(\pm)}_{\un a}= \partial_{\un a} X \pm \epsilon_{{\un a}{\un b}{\un c}}\partial^{\un b} X^{\un c} .
\end{align}
The momenta conjugate to $X$ and $X^a$ are
\begin{align}
{P} = {\dt X}
~~~\mathrm{and}~~~
{P}_a = \dt X_a + \partial_a X^0
,
\end{align}
so that the action becomes (\ref{E:HamiltonianAction}) with Hamiltonian
\begin{align}
H = \int \left[ 
	\tfrac12 {P}^2 
	+\tfrac12 {P}_a^2 
	+ \tfrac12 (\partial_{a} X)^2
	+\tfrac14 F_{{a}{b}}^2 
	+ X^0 \partial^{a} {P}_{a}
	\right] \, d^2\sigma .
\end{align}
The $\tau$ component $X^0$ of the gauge 1-form is the Lagrange multiplier for the Gau\ss{} law constraint 
\begin{align}
\U = \partial^{a} {P}_{a} .
\end{align}
The 
$\mathrm G = \gl2R$-covariant 
currents
\begin{align}
P := F^{(+)}_\tau 
	= {P} + \epsilon_{{a}{b}} \partial^{a} X^{b}
~~~\mathrm{and}~~~
P_{a} := \epsilon_{0 {a}}{}^{b}  F^{(+)}_{b}
	= {P}_{a} + \epsilon_{{a}{b}} \partial^{b} X
\end{align}
commute with the Gau\ss{} law constraint $\mathcal U$. Their quantum bracket algebra is 
\begin{align}
[P , P ] =0
~,~~
[P , P_{a} ] = 2i \epsilon_{{a}{b}} \partial^{b} \delta
~,~~
[P_{a} , P_{b} ]=0 .
\end{align}

\subsection{2D String}
\label{limbo.2}

The 2D bosonic string arises as the theory of a pair of 1-forms $X_{{\un a}i}$ ($i=1,2$) with fieldstrengths $F_{{\un a}{\un b} i} = \partial_{[{\un a}}X_{{\un b}]i}$ on a worldvolume locally of the form $\mathbf R^{2,2}$. We define the (anti-) selfdual combinations (indices are raised and lowered with $\delta_{ij}$)
\begin{align}
F^{(\pm)}_{{\un a}{\un b}i} := F_{{\un a}{\un b} i} \pm \tfrac12 C_{ij}\epsilon_{{\un a}{\un b}}{}^{{\un c}{\un d}} F_{{\un c}{\un d}}^{\, j}
\end{align}
in terms of which the action has SO(2,2)$\times$SO(2) symmetry. The momentum conjugate to $X_i$ is given by 
\begin{align}
{P}_{{a} i} = \dt X_{{a} i} + C_{ij} \epsilon_{{a}{b}{c}}\partial^{b} X^{{c} j}. 
\end{align}
This gives the $\mathrm H = \sl2R \times \so2$-invariant Hamiltonian action
\begin{align}
H &= \int \left[ \tfrac12 {P}_{{a} i}^2 + \tfrac14 F_{{a}{b} i}^2 + X^{0 i} \partial^{a} {P}_{{a}i} \right] \, d^4\sigma.
\end{align}
The $X^{0 i}$ component acts as a Lagrange multiplier for the Gau\ss{} law constraint
\begin{align}
\U_i = \partial^{a} {P}_{{a}i}.
\end{align}
The covariant derivatives become
\begin{align}
P_{{a}i}:=F^{(+)}_{0 {a}i} = {P}_{{a}i}  + C_{ij} \epsilon_{{a}{b}{c}} \partial^{b} X^{{c} j} ,
\end{align}
and their quantum brackets are ($\eta_{{a}i{b}j{c}} = C_{ij} \epsilon_{{a}{b}{c}}$)
\begin{align}
[P_{{a} i}, P_{{b} j}] = 2i C_{ij} \epsilon_{{a}{b}{c}} \partial^{c} \delta .
\end{align}
Again, we see that the H symmetry is enhanced to $\mathrm G= \sl3R \times \sl2R$ in the current algebra. 
This system results from the double dimensional reduction \cite{Linch:2015qva} of the 5-brane theory \cite{Linch:2015fya} wrapped on a 2-torus:
Starting with $\mathrm G = \mathrm{SL}(5; \mathbf R)$, the double reduction gives
$\partial^a \to (\partial^a , \partial^i) \to (\partial^a, 0)$ and 
$P_{ab} \to (P_{ab}, P_{ai}, C_{ij}P^+ ) \to (0, P_{ai},0)$.

\section{Fermions and Internals}

The scalars of extended supergravities in various dimensions have long been known to live in various coset spaces, including exceptional groups as the symmetry groups in the higher-rank cases \cite{Cremmer:1979up,Cremmer:1978ds,Julia:1980gr}.  One suggestive way to incorporate this structure is to introduce the momenta of extra dimensions as central charges gauged in the covariant derivatives \cite{Siegel:1980bp}, as would be implied by dimensional reduction from a supergravity theory in dimensions higher than that of string (``S-")theory, M-theory, or even the original concept of F-theory.  We'll refer to such much-higher-dimensional embeddings of S-theory as ``F-theory", and the massless sectors as ``F-gravity" (or ``F-supergravity" when including fermions).  F-gravity has been described both with momenta vanishing in the extra embedding dimensions \cite{West:2001as, Hull:2007zu, Berman:2010is} and with these momenta nonvanishing before application of ``section conditions" \cite{Coimbra:2011ky,Berman:2012vc}, as a generalization of the doubled-dimension method applied to manifest T-duality by embedding S-theory in ``T-theory" \cite{Siegel:1993xq,Siegel:1993th,Siegel:1993bj}.  Often the extra dimensions are used just to incorporate the symmetries of the scalars (and differential forms); here we concentrate on the ``dual" approach where the coset space includes the metric; in particular, this means the isotropy group is the maximally noncompact (``split") version, which modifies reality properties (essentially by Wick rotation):  The resulting sectioning thus includes dimensional reduction in a timelike dimension \cite{Hull:1998br}.  (E.g., to relate our D=6 equations to the expressions of \cite{Siegel:1980bp} for D=4, the central-charge partial and covariant derivatives should be switched with those for spacetime.)

We now generalize our previous results for bosonic sectors of F-theories in lower dimensions (D=3 and 4) \cite{Linch:2015fya,Linch:2015qva} by (1) including examples in even lower dimensions (described above) for completeness, (2) including D$'$ ``internal" coordinates $Y^{a'}$ so that all these examples represent the critical dimension D+D$'$=10, but with varying (D-dimensional) degrees of U duality manifest, and (3) adding the fermionic coordinates for supersymmetry.
In earlier papers \cite{Linch:2015lwa,Linch:2015fya,Linch:2015qva} we ignored ``internal" dimensions \cite{Hohm:2013pua,Hohm:2013vpa,Hohm:2013uia,Godazgar:2014nqa,Hohm:2014fxa,Musaev:2014lna}, additional to those required to construct the coset space for F-theory embeddings of classical superstring theories in lower dimensions D=3,4,6.  Here we generalize to S-theory in 10 dimensions, including 32 supersymmetries via Green-Schwarz coordinates $Î^{\un \alpha}$, and adding the missing spacetime coordinates $Y$ in an ``ordinary string way" as scalars on the worldvolume, in contrast with the spacetime coordinates $X$ implied by the coset construction, which appear as worldvolume forms with selfdual field strengths.  
This allows us to consider all F-gravities treated in the literature, with ``spacetime" dimension D=1 through 6, and internal dimension D$'$=10$-$D.  (There is a supersymmetry obstruction to considering the case D=7, related to D$'$=3 supergravity having no vectors when its maximal symmetry is manifest:  See the Prospects in section \ref{Conclusions}.  Some progress has been made in constructing at least the bosonic sector of the D=10 theory with E$_{11}$ symmetry \cite{West:2001as,West:2003fc}.)  

In the original subcritical bosonic construction, T-duality was used for the field strengths of the worldvolume forms $X$ to select selfdual currents defining the theory.
In the present case scalars are not dual to (pseudo)scalars, so we introduce form gauge fields $\widetilde Y$ dual to scalars $Y$ (in the sense of Hodge duality of their field strengths), and impose
selfduality of this combination \cite{Cremmer:1998px}.  
(For S-theory, where $\widetilde Y$ are pseudoscalars, this doubling for selfduality in the ordinary bosonic string action was introduced in \cite{Tseytlin:1990nb,Tseytlin:1990va}.)
Afterwards the terms for the dual internal coordinates $\widetilde Y$ can be dropped, or the $Y,\widetilde Y$ combination can be kept and treated in the same way as the other coordinates $X$, using only the selfdual currents.

The symmetry structure is summarized in table \ref{T:structure}.
\begin{table}[ht]
$$ \vcenter{
\halign{ #â & #â & #â & #â & #â & #â & #â & #â & #â & #â & # \cr
D & d & L & H & G & $P$ & $\S$ & $\U$ & $\V$ & $\W$ \cr
1 & 3 & SL(2) & GL(1) & GL(2) & 2+1 & 2 & 1 & & \cr
2 & 4 & SL(2)$^2$ & GL(2) & SL(3)SL(2) & (3,2) & (3$'$,1) & (1,2) & & \cr
3 & 6 & GL(4) & Sp(4) & SL(5) & 10 & 5$'$ & 5 & & \cr
4 & 11 & \gl4C & \sp4C & SO(5,5) & 16 & 10 & 16$'$ & 1 & \cr
5 & 28 & SU*(8) & USp(4,4) & E$_{6(6)}$ & 27 & 27$'$ & 78+1 & 27 & \cr
6 & 134 & SU*(8)$^2$ & SU$*$(8) & E$_{7(7)}$ & 56 & 133 & 912 & 133 & 1 \cr
}
} \hskip-.25in $$
\begin{caption}{Symmetry structure of F-theories making up the D-dimensional S-theories}
\label{T:structure}
\end{caption}
\end{table}
(GL(1) charges are not listed.)  The D-dimensional metric and associated bosons live in the coset space G/H.  (Note that H$_{\rm D}$ with half the argument = covering group of SO(D$-$1,1).  The doubled argument necessitates Type II, unless modified by branes/boundaries.)  The remaining ``scalars" live in G$'$/H$'$, where G$'_{\rm D}$=GL(D$'$) and H$'_{\rm D}$=SO(D$'$).  The internal coordinates $Y(§)$ are in the defining (vector) representation of the internal symmetry H$'_{\rm D}$.  The remaining ``vectors" couple to $Y$-momenta as ``charges" (but we treat general dependence on $X$ and $Y$ coordinates).  The table lists the dimensions of the G representations of the momenta $P(§)$, dual to coordinates $X(§)$.  (In the dual analog of maximal supergravity in D$'$ dimensions, the coset bosons of G/H correspond to the scalars, the metric appears as the bosons of the coset G$'$/H$'$, and the vectors couple to $P$ as central charges.)  Table \ref{T:structure} also contains the G representations of the constraints $\S,\U,\V,\W$ \cite{Linch:2015fya, Linch:2015qva}. (The $\S$ constraint acts as the worldvolume derivatives $»/»§$ with respect to the coordinates $§$, less $ $.  The $\W$ constraint is new, and is described in section \ref{S:Constraints}, along with the rest.)  L is the manifest symmetry of the action, which is closely related to the dimension d of the worldvolume. 

\section{Indices}


There is a profusion of index types, so we give a catalog here.  
Besides the G/H coset space there is also the internal (group) space H$'$.  For the most part we'll deal with supersymmetry, and thus fermions, and so will use primarily spinor indices for the groups H and H$'$:  
These will be (lower-case) Greek, respectively unprimed and primed.
(Bars and dots, as well as raising and lowering, also appear when spinor representations aren't unique.  Here dots generally refer to complex conjugate representations, while bars just mean independent ones.)

However, for some general remarks (and when $©$ matrices are necessary) we'll need vector indices, and super indices (calligraphic Roman):  For the former we use upper-case Roman letters, with a prime for H$'$.  (In particular, H vector indices will be used for discussing G representations.)  We also use underlined lower-case Greek for combined (H$ð$H$'$) spinor indices for the fermions.  (These are the fusion of 2 spinor indices, with reduction in the vector cases.) 
Also, although in spinor notation the worldvolume coordinates also carry H spinor indices, for purposes of general discussion we'll use lower-case Roman letters.  


Thus, ignoring (a) the distinction on spinor indices from raising/lowering, bars, and dots, and (b) reduction (by symmetrization, traces, etc.) on bispinor vector indices, we have:

$$ \vcenter{
\halign{ #â & #â & $#$ \cr
\bf representation & \bf group & \hbox{\bf index} \cr
spinor & H & Π\cr
& H$'$ & \alpha' \cr
& H$ð$H$'$ & \un Œ = Œ\alpha' \cr
$§$-vector & H & a = \alpha \beta \cr
vector & H & A = μ \cr
& H$'$ & A' =\un a a'= μ\alpha'\beta' \cr
& H$ð$H$'$ & {\un A}\, = A,A' = A, a', aa' \cr
super & H$ð$H$'$ & \A = \un Œ,{\un A},a\un Œ \cr
}
} $$
\vskip.2in

\noindent Symmetrization $(¼)$, antisymmetrization $[¼]$, and their graded analogs $(¼]$ and $[¼)$, have coefficients $à1$ for each term.  ($|¼Ê|$ indicates omission of indices from the (anti)symmetrization.)

\section{General}

The graded quantum brackets of the current algebras are a generalization of (one handedness of) those in string theory:
\begin{align}
[ \dd_{\A} (1), \dd_{\B} (2) Õ &= 
	f_{{\A} {\B}}{}^{\C} \dd_{\C} \, \delta (1-2)
	+2i \eta_{{\A}{\B} c} \, \partial^c \delta (1-2) 
	. 
\end{align}
with $»^a­»/»§_a$ and $¶=¶^{\rm d-1}(§)$,
in terms of the structure constants $f$ and the generalized metric $ú$.  
(String results d=2 can be obtained here and in the following by choosing ``$a$" to take only 1 value.)
Because of selfduality, ``time" and ``space" components of the currents need not be treated separately.
This ``metric" is essentially Clebsch-Gordan-Wigner (CGW) coefficients relating (adjoint$°$adjoint)$_{sym}$ to the representation of $»^a$; it has no relation to the Cartan metric, which vanishes in flat superspace.
For lower D, where $X,Y,\widetilde Y$ are worldvolume differential forms, it is the Levi-Civita tensor $·$ for the worldvolume; otherwise it gives a generalization of Hodge duality.

The Jacobi identities for the currents imply the usual Lie algebra Bianchi identity
\begin{equation}
\label{Lie}
f_{[{\A}{\B}|}{}^{\D} f_{{\D}|{\C})}{}^{\E} = 0
\end{equation}
as well as a generalization of the (graded) antisymmetry condition of 2D current algebra (and simple Lie algebras)
\begin{equation}
\label{anti}
f_{{\A}({\B}|}{}^{\D} ú_{{\D}|{\C}]c} = 0 .
\end{equation}
We also have in general at least the part of the Virasoro constraints corresponding to $§$ (less $ $) derivatives
\begin{align}
\label{E:Section}
\mathcal S^{c} = \f14 ú^{{\A}{\B} c}\dd_{\A} \dd_{\B} .
\end{align}

The next step is to separate currents by (spacetime-engineering) dimension:  The algebra is a direct generalization of the result of  \cite{Siegel:1985ra,Siegel:1985xj}.  Then all bosonic currents are lumped together as $\P_{\un A}$, while the fermionic ones are $D_{\un \alpha}$ and its ``dual" $¯^{a {\un \alpha}}$:
\begin{subequations}
\begin{align}
\label{DD}
ÓD_{\un \alpha},D_{\un \beta}Õ & = 2(©^{\un A})_{{\un \alpha}{\un \beta}}\P_{\un A}ʶ \\
[D_{\un \alpha},\P_{\un A}] & = 2ú_{{\un A}{\un B}c}(©^{\un B})_{{\un \alpha}{\un \beta}}¯^{a {\un \beta}}ʶ\\
& \nonumber \\
\label{Abel}
[\P_{\un A},\P_{\un B}] & = 2iú_{{\un A}{\un B}c}Ê»^c ¶ \\
ÓD_{\un \alpha},¯^{{a} {\un \beta}}Õ & = 2i¶_{\un \alpha}^{\un \beta}Ê»^a ¶ .
\end{align}
\end{subequations}
$ÓD,DÕ$ defines the supersymmetry transformations on the coordinates
\begin{equation}
ÓD_{\un \alpha},D_{\un \beta}Õ = 2(©^{\un A})_{{\un \alpha}{\un \beta}}\P_{\un A}âÛâ¶Î^{\un \alpha} = ·^{\un \alpha}¼,â¶{X}^{\un A} = i·^{\un \alpha}(©^{\un A})_{{\un \alpha}{\un \beta}} Î^{\un \beta} .
\end{equation}
The $f$ ``$ú©$" follows from the generalized $©$-matrices $f_{{\un \alpha}{\un \beta}}{}^{\un A}=2(©^{\un A})_{{\un \alpha}{\un \beta}}$ in $ÓD,DÕ$ by use of the $fú$ antisymmetry Bianchi identity (\ref{anti}), the nonvanishing parts of the metric $ú_{{\un A}{\un B}{c}}$ and 
\begin{equation}
ú_{\un \alpha}{}^{{b} {\un \beta}}{}_{c} = ¶_{\un \alpha}^{\un \beta} ¶_{c}^{b} .
\end{equation}
There is also the nontrivial $ff$ Lie identity (\ref{Lie})
\begin{align}
\label{Jacob}
[ÓD_{(\un \alpha} , D_{\un \beta}Õ , D_{\un \gamma)}]^¯  = 0
\end{align}
of S-theory and super Yang-Mills, which implies the Fierz identity
\begin{align}
\label{Fierz}
\eta_{{\un A}{\un B}{c}} (\gamma^{\un A})_{(\un \alpha \un \beta} (\gamma^{\un B})_{\un \gamma)\un \delta} 
= 0 .
\end{align}
To absorb naked indices, it is sometimes more convenient to use the form
\begin{equation}
\label{Chubby}
ú_{{\un A}{\un B}{c}}(©^{\un A} Â)(©^{\un B} Â) = 0
\end{equation}
where $Â^{\un \alpha}$ is an arbitrary bosonic spinor/``twistor" and the last $Â$ is gratuitous.

The explicit dependence of this algebra on fermionic coordinates $Î$ is
\begin{subequations}
\begin{align}
D_{\un \alpha} &= \Pi_{\un \alpha} + (ß\P_{\un A} -iú_{{\un A}{\un B}{c}}Ω^{\un B} »^c Î)(©^{\un A} Î)_{\un \alpha} \\
\P_{\un A} &= ß\P_{\un A} -2iú_{{\un A}{\un B}{c}}Ω^{\un B} »^c Î \\
¯^{{a} {\un \alpha}} &= -2i»^a Î^{\un \alpha}
\end{align}
\end{subequations}
where 
\begin{equation}
Ó\Pi_{\un \alpha},Î^{\un \beta}Õ = ¶_{\un \alpha}^{\un \beta} ¶ .
\end{equation}
For $ÓD,DÕ$ to work, the Fierz identity (\ref{Fierz}) must again be satisfied.
Here $ß\P$ is a representation of the Abelian, bosonic subalgebra (\ref{Abel}).
(It commutes with $Î^{\un \alpha}$ and $\Pi_{\un \alpha}$.)
In terms of coordinates ${X}^{\un A}$, these currents appear in the manifestly selfdual combination
\begin{equation}
ß\P_{\un A} = {P}_{\un A} + ú_{{\un A}{\un B}{c}}»^c{X}^{\un B}
\end{equation}
(effectively ${P}+\ast dX$), where
\begin{equation}
[ {P}_{\un A} , {X}^{\un B} ] = -i¶_{\un A}^{\un B} ¶ .
\end{equation}

Another useful representation, which eliminates the $Î^2»Î$ term in $D$, is
\begin{subequations}
\begin{align}
\label{rap}
D_{\un \alpha} &= \Pi_{\un \alpha} + {P}_{\un A} (©^{\un A} Î)_{\un \alpha} - 2{X}^{\un A} ú_{{\un A}{\un B}{c}}(©^{\un B} »^c Î)_{\un \alpha} \\
\P_{\un A} &= {P}_{\un A} + ú_{{\un A}{\un B}{c}}(»^c {X}^{\un B} -iΩ^{\un B} »^a Î) \\
¯^{{a} {\un \alpha}} &= -2i»^a Î^{\un \alpha} .
\end{align}
\end{subequations}
This is a convenient starting point to generate others by the transformation
\begin{equation}
\dd\mapsto e^{-\O}\dd e^\O¼,â\O := cÇ{X}^{\un A} ú_{{\un A}{\un B}{c}}Ω^{\un B} »^c Î ;
\end{equation}
\begin{align}
D_{\un \alpha} &= \Pi_{\un \alpha} + [{P}_{\un A} + cú_{{\un A}{\un B}{c}} (»^c {X}^{\un B} -iΩ^{\un B} »^c Î)](©^{\un A} Î)_{\un \alpha} - 2(1-c) {X}^{\un A} ú_{{\un A}{\un B}{c}}(©^{\un B} »^c Î)_{\un \alpha} \cr
\P_{\un A} &= {P}_{\un A} + ú_{{\un A}{\un B}{c}}[»^c {X}^{\un B} -i(1+c)Ω^{\un B} »^c Î] \cr
¯^{{a} {\un \alpha}} &= -2i»^a Î^{\un \alpha} .
\end{align}
Supersymmetry and translations in these representations are
\begin{align}
q_{\un \alpha} &= ÇÓ {\Pi}_{\un \alpha} 
	-(©^{\un A} Î)_{\un \alpha} [ {P}_{\un A} +ú_{{\un A}{\un B}{c}} ( c»^c{X}^{\un B} -i(c-\f23)Ω^{\un B} »^c Î )] Õ \cr
p_{\un A} &= Ç[{P}_{\un A} +i(1-c)ú_{{\un A}{\un B}{c}}Ω^{\un B} »^c Î] .
\end{align}
Notable examples are $c=1$ for our original case, and $c=-1$ and $\f23$ to simplify $P$ and $q$.

The fermions drop out of the Virasoro (\ref{E:Section}) constraints after applying $D_{\un \alpha}=0$:
\begin{equation}
\S^{c} = \f14 ú^{{\un A}{\un B}{c}}\P_{\un A}\P_{\un B} ,
\end{equation}
where we have defined the dual $ú^{{\un A}{\un B}{c}}$ to $ú_{{\un A}{\un B}{c}}$. A duality-covariant solution to the corresponding section condition (at least for one momentum with itself) is provided by the twistors introduced in (\ref{Chubby}):\begin{equation}
ú^{{\un A}{\un B}{c}}p_{\un A} p_{\un B} = 0âÜâp_{\un A} = ©_{\un A} Â
\end{equation}
modulo ``pure spinors" $©_{\un A} Â$ = 0, assuming the existence of a $©_{\un A}$ that satisfies the same Fierz as $©^{\un A}$.
We can also define a $(©^{\un A})^{{\un \alpha}{\un \beta}}$ that combines with $(©^{\un A})_{{\un \alpha}{\un \beta}}$ to form generalized Dirac-matrices that satisfy a generalization of a Dirac/Clifford algebra.
We will be more explicit about this algebra presently.

\section{Branes}

We now separate $X^{\un A}=(X^A,Y^{A'})$ and $\P_{\un A}=(\P_A,\P_{A'})$. 
Here $Y^{A'} = (Y^{a'},\widetilde Y_{aa'})$ are the scalars and their duals and $\P_{A'}=(ç_{a'},\di^{aa'})$ are the Ramond-Ramond currents $\Upsilon_{a'}$ (for $Y$) with partner $\digamma^{aa'}$ (for $\widetilde Y$) \cite{Hatsuda:2014aza}. 
The algebra then expands out to 
\begin{subequations}
\begin{align}
Ó D_{\un \alpha} , D_{\un \beta} Õ &= 
	2(\gamma^A)_{{\un \alpha} {\un \beta}} \P_A \, \delta 
	+2(\gamma^{a'})_{{\un \gamma}({\un \alpha}}
		(\mho_{a'})_{\un Ã)}{}^{\un ©}
		\, \delta  \\
[ D_{\un \alpha} , P_A ] &= 2 \eta_{ABc} (\gamma^B)_{{\un \alpha} {\un \beta}} \Omega^{c {\un \beta}} \, \delta \\
[ D_{\un \alpha} , \Upsilon_{a^\prime }  ] &= 2 (\Gamma_b\gamma_{a'})_{({\un \alpha} {\un \beta})}
	\Omega^{b{\un \beta}} \, \delta \\
[ D_{\un \alpha} , \digamma^{aa^\prime } ] &= 2 (\gamma^{a'})_{{\un \alpha} {\un \beta}} \Omega^{a{\un \beta}} \, \delta \\
& \nonumber \\
\label{E:PP}
[ P_A , P_B ] &= 2i \eta_{ABc} \, \partial^{c} \delta\\
Ó D_{\un \alpha} , \Omega^{b{\un \beta}} Õ &= 2i \delta_{\un \alpha}^{\un \beta} \, \partial^b\delta  \\
[ \mho_{a^\prime}, \mho_{b^\prime}] &= 2i \eta_{a^\prime b^\prime}Ö»\deltaââ
( [ \Upsilon_{a'} , \digamma^{bb'} ] = 2i ¶_{a'}^{b'} \, \partial^b \delta ) .
\end{align}
\end{subequations}
where $\Upsilon$ and $\digamma$ have been combined into
\begin{align}
\label{Larry}
\mho_{a'} = \Upsilon_{a'} I + \digamma^a{}_{a'} \Gamma_a
\end{align}
corresponding to
\begin{equation}
¶_{\un \alpha}^{\un \beta} (ÀY_{a'}+»^a \widetilde Y_{aa'}) + (\Gamma_a)_{\un \alpha}{}^{\un \beta} (À{\widetilde Y}{}^a{}_{a'} + »^a Y_{a'})
\end{equation}
in the Lagrangian.
We also have
\begin{equation}
ú_{a'}{}^{bb'}{}_c = ¶_c^b ¶_{a'}^{b'} .
\end{equation}

While $©_{a'}$ are the usual vector $©$-matrices for SO(D$'$), $©_A$ and $\Gamma_a$ are more general CGW coefficients for more general groups. 
They are related by an Dirac/Clifford-like equation
\begin{align}
(\gamma^A)_{{\un \alpha} {\un \gamma}}(\gamma^B)^{{\un \gamma}{\un \beta}} 
	+ (\gamma^B)_{{\un \alpha} {\un \gamma}}(\gamma^A)^{{\un \gamma}{\un \beta}}=2\eta^{AB\un c} (\Gamma_{\un c})_{\un \alpha}{}^{\un \beta}
.
\end{align}
Equivalently, we can define 
\begin{align}
(\Gamma_0)_{\un \alpha}{}^{\un \beta} = \delta_{\un \alpha}{}^{\un \beta}
~~~\textrm{and}~~~
(\Gamma_c)_{\un \alpha}{}^{\un \beta} = ü©^A_{{\un \alpha}{\un \gamma}}©^{B{\un \gamma}{\un \beta}}ú_{ABc} .
\end{align}
(As described earlier, the spinor space is a direct product of a worldvolume space on which $©^A$ and $\Gamma_a$ act, and an internal space acted upon by $©^{a'}$.)
The Fierz identity (\ref{Fierz}) is now
\begin{align}
\eta_{ABc} (\gamma^A)_{({\un \alpha} {\un \beta}} (\gamma^B)_{{\un \gamma})\un \delta} 
	+ (\Gamma_c\gamma_{a'})_{({\un \alpha} {\un \beta}} (\gamma^{a'})_{{\un \gamma})\un \delta}
	+ (\gamma^{a'})_{({\un \alpha} {\un \beta}} (\Gamma_c\gamma_{a'})_{{\un \gamma})\un \delta}
= 0 .
\end{align}
Due to the mixed first/second-class constraints $D=0$, the $\S$ constraint reduces to
\begin{align}
\label{E:S}
\mathcal S^c =
	\f14 \eta^{AB c} P_A P_B
	+ ü\digamma^{cc^\prime} \Upsilon_{c'}  .
\end{align}

\section{Constraints}
\label{S:Constraints}

First-quantized relativistic theories (particles, strings, etc.)¼are completely defined by their constraints.
The constraints $\U$, $\V$, \dots (and the corresponding dimensional reduction and section conditions obtained by replacing currents with 0-modes) can be found from $\S$ (\ref{E:S}) by the method described in \cite{Linch:2015qva}.
(They can also be derived from the action:  For example, $\U$ is Gauss's law.)
To obtain $\U$ we take the difference 
\begin{align}
\label{E:Uconstraint}
\S^a - \tilde \S^a
	= (\partial^a X^{\un A}) \P_{\un A} 
	+ \mathcal O(\U). 
\end{align}
The left-hand side of this equation gives the combination $\eta^{{\un C} {\un A} {a}} \eta_{{\un C}{\un B} {b}} (\partial^{{b}} X^{\un B})\P_{\un A}$ for the $X$ term whereas we require the first term on the right-hand side to be $\delta^{\un A}_{\un B} \delta^{{a}}_{{b}}$. Defining the difference
\begin{align}
U^{{\un A} {a}}_{{\un B} {b}} :=
	\delta^{\un A}_{\un B} \delta^{{a}}_{{b}}
	- \eta^{{\un C} {\un A} {a}} \eta_{{\un C}{\un B} {b}} ,
\end{align}
we find that the constraint is
\begin{align}
\U^{{a}}_{\un B} := U^{{\un A} {a}}_{{\un B} {b}} \, \partial^{{b}} P_{\un A} .
\end{align}
(For example in three dimensions $U^{A c}_{B d} = U^{[{a_1}{a_2}] {c}}_{[{b_1}{b_2}]d} \propto \delta^{[{a_1}}_d\delta^{{a_2}]}_{[{b_1}}\delta^{{c}}_{{b_2}]}$ so that $\U^{{a}}_B = \U^{{a}}_{[{b_1}{b_2}]} = \delta^{{a}}_{[{b_1}} \partial^{{c}} P_{{b_2}] {c}}$ which is equivalent to $\U_{{a}} = \partial^{b} P_{[{a} b]}$ \cite{Linch:2015fya}.
Similarly, we obtain for the $Y$ terms the tensor 
$U_{aa'\, c}^{bb'\, d} = \delta_{a'}^{b'} \delta_a^{[b}\delta_c^{d]}$
and 
$\U^{b\, aa'} = \partial^{[b} P^{a]a'}$;
in a Hamiltonian action, $\widetilde Y_{abb'}$ is the Lagrange multiplier.)

At this point, we have that $\U^{{a}}_{\un B} = \tfrac12 U^{{\un A} {a}}_{{\un B} {b}} \, \partial^{{b}} (\dd_{\un A} + \tilde \dd_{\un A})$. We can then define $\V$ by $ U \partial(\dd - \tilde \dd) = \mathcal O(\V)$. The left-hand side evaluates to 
$
U^{{\un A} {a}}_{{\un B} {b}} \, \partial^{{b}} (\dd_{\un A} - \tilde \dd_{\un A})
	= U^{{\un A} {a}}_{{\un B} {b}}  \, \eta_{{\un A} {\un C} {c}} \partial^{{b}} \partial^{{c}} X^{\un C}
$ so that
\begin{align}
\V^{{a}}_{{\un A} {\un B}} 
	:= \tfrac12 V^{{a}}_{{\un A}{\un B} \, cd} \,\partial^c \partial^d
~~~&\mathrm{with}~~~
V^{{a}}_{{\un A}{\un B} \, cd } := U^{{\un C} {a}}_{{\un A} (c} \, \eta_{{\un C} {\un B}d)}
.
\end{align}
Expanding this out, we see that $\V^{{a}}_{{\un A}{\un B}}
= \V^{{a}}_{{\un B}{\un A}}$ is symmetric. 
(For D $\leq 3$, there are no invariant tensors in the proper representation of G so that in this case $\V=0$. For D=4, $\mathrm G= \mathrm{Spin}(5,5)$ so that 
$V^e_{AB\, cd} = (\Gamma^e)_{AB}\eta_{cd}$ giving $\V^e_{AB} = (\Gamma^e)_{AB} \V$ with singlet $\V = \tfrac12 \eta_{cd} \partial^c \partial^d$.)

Finally, there is another constraint for D $\geq 6$. To see this, we recall that the requirement (\ref{E:Uconstraint}) arises from the closure of the C-bracket since 
\begin{align}
\label{E:C-bracket}
[ V_1^A \dd_A , V_2^B \dd_B ] 
	&= 2i \eta_{ABc} \partial^c \delta \,  V_1^AV_2^B
-i \delta \, \left[ V_{[1}^C\partial_C V_{2]}^D
		-\tfrac12 \eta_{ABe} \eta^{CDe}V_{[1}^A\partial_C V_{2]}^B \right] \dd_D 
\end{align}
only up to $\mathcal U$ constraint (and second class constraints). The truncation of the C-bracket (\ref{E:C-bracket}) to massless fields results in the exceptional-and-``doubled'' geometry bracket. 
Thus, the massless truncation of this structure must reproduce the geometry associated to the homogeneous spaces $\mathrm E_{n(n)}/\mathrm H_n$ for each rank $n$. In particular the C- and D- brackets should reduce to the exceptional Courant and Dorfman brackets worked out in references \cite{Coimbra:2011ky, Berman:2012vc}.

Matching up with the $Y$ tensor of reference \cite{Berman:2012vc} requires that it be equal to
$\eta_{ABe}\eta^{CDe}$ up to terms that vanish by constraints. Thus $\U$ suffices for D$²$5 (as we have found above) but for $\mathrm E_{7(7)}$, there is an additional term of the form $C_{AB}C^{CD}$ in the $Y$ tensor where $C$ is the $\mathrm{Sp}(56; \mathbf R)$ invariant. This vanishes provided $C^{AB} \partial_A f P_B$ vanishes modulo constraints. 
This, in turn, is implied by the existence of the singlet constraint
\begin{align}
\W := C^{AB} \dd_A \tilde \dd_B. 
\end{align}

The various constraints can then be applied to section both spacetime and the worldvolume to reduce F-theory to M, T, and S-theory:

$$
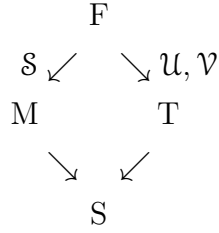
\vcenter{\halign{#&¼#¼&#\cr
	& F&\cr
	¼$\S\swarrow$&&$\searrow\U,\V$\cr
	M\hfil&& \hfil T\hfil\cr
	¼$\searrow$&&$\swarrow$\cr
	& S&\cr}}
 $$
~\\
\noindent 
For $\mathrm D < 6$, the constraints take the form (cf.\ ref.\ \cite{Linch:2015qva})
\begin{align}
&~\hbox{Virasoro} &
	\mathcal S^c =¼& \tfrac14 \eta^{{\un A} {\un B} c} \P_{\un A} \P_{\un B} && \cr
&\begin{array}{l} \hbox{dimensional} \\ \hbox{reduction}\end{array} & 
	\on\circ\S{}^c :=¼& \eta^{{\un A} {\un B} c} p_{\un A} \P_{\un B} & 
	\U^{{a}}_{\un B} = ¼& U^{{\un A} {a}}_{{\un B} {b}} \, \partial^{{b}} P_{\un A}   ¼&  &\cr
& \begin{array}{l} \hbox{section} \\ \hbox{condition}\end{array} & 
	\under\circ\S{}^c :=¼& \eta^{{\un A} {\un B} c} p_{\un A} p_{\un B} & 
	{\under\circ\U}{}^{{a}}_{\un B} := ¼& U^{{\un A} {a}}_{{\un B} {b}} \, \partial^{{b}} p_{\un A} 
 & 
\V^{{a}}_{{\un A} {\un B}} 
	=  ¼& V^{{a}}_{{\un A}{\un B} \, cd} \,\partial^c \partial^d 
\nonumber
\end{align}
where $p$ is the 0-mode of ${P}$. 
For $D\geq 6$, the $\W$ constraint (and possibly others) must be incorporated.
We leave the analysis of these higher-dimensional cases for future work. 
 
\section{Cases}
\label{Casey}

Fermionic currents $D$ are in representations of H$ð$H$'$; bosonic currents $P$ are in representations of H, $\Upsilon$ in representations of H$'$; they are all spinor$°$spinor:

$$ \vcenter{
\halign{ #¼ & #¼ & #â & #â & #â & #â & #â & #â & # \cr
D & d & L & H & H$'$ & $D$ & $P$ & $\Upsilon$ & $»$ \cr
1 & 3 & SL(2) & GL(1) & SO(9) & 16+16 & 1+1+1 & 9 & 1+1 \cr
2 & 4 & SL(2)$^2$ & GL(2) & SO(8) & (2,8)+(2,8$'$) & 3+3 & 8$_v$ & 3 \cr
3 & 6 & GL(4) & Sp(4) & SO(7) & (4,8) & 10 & 7 & 5 \cr
4 & 11 & \gl4C & \sp4C & SU(4) & (4,4)+($Ð4$,$Ð4$) & 4$°Ð4$ & 6 & 5+Ð5 \cr
5 & 28 & SU$*$(8) & USp(4,4) & USp(4) & (8,4) & 27 & 5 & 27 \cr
6 & 134 & SU$*$(8)$^2$ & SU$*$(8) & SU(2)$^2$ & (8;2,1)+(8$'$;1,2) & 28+28$'$ & (2,2) & 70+63 \cr
}
} \hskip-.1in $$
\vskip.2in
\noindent (The bosonic current $\di$ is partly $ß\di = »Y$, so in the H representation of $»$ and the H$'$ vector representation.  Similarly, the fermionic current $¯=-i»Î$ is in the unreduced representation of $»$ $°$ the dual to $D$.)  
In index notation:

$$ \vcenter{
\halign{ #â & $#$â & $#$â & $#$â & $#$ \cr
D & D_{\un Œ} & P_A & \Upsilon_{a'} & »^a \cr
1 & _{à\alpha'} & _+Ê,¼_-Ê,¼_0 & _{(Œ'º')} & _+Ê,\, _- \cr
2 & _{Œ\alpha'}Ê,¼_{Œ{\bar \alpha}'} & _{(Œº)}Ê,{\overline{~}}_{(\alpha \beta)} & _{{\alpha}'{\bar \beta}'} & _{(Œº)} \cr
3 & _{\alpha\alpha'} & _{(Œº)} & _{[\alpha'\beta']} & _{ҌºÔ} \cr
4 & _{\alpha\alpha'}Ê,~{}_{ÀŒ}{}^{\alpha'} & _{ŒÀº} & _{[\alpha'\beta']} & _{ҌºÔ}Ê,¼_{ÒÀŒÀºÔ} \cr
5 & _{\alpha\alpha'} & _{ҌºÔ} & _{Ò\alpha'\beta'Ô} & _{ҌºÔ} \cr
6 & _{\alpha\alpha'}Ê,¼^{Œ{\bar \alpha}'} & _{[Œº]}Ê,¼^{[Œº]} & _{\alpha'}{}^{{\bar \beta}'} & ^{[Œº©¶]}Ê,¼_Œ{}^º \cr
}
} $$
\vskip.2in

\noindent Primed Greek indices are (irreducible) SO(D$'$) = SO(10$-$D) spinor indices.  Unprimed Greek indices are the same for H$_{\rm D}$, which are the same as for SO(D$-$1,1) except for double the range (see previous table); thus there is some symmetry between $P_D$ and $\Upsilon_{10-D}$.  $©$ matrices and $ú$'s (H) are all Kronecker $¶$'s since spinor notation is used for $P$; $©'$ matrices are also except for $\mathrm D<4$ ($\mathrm D'>6$), where the usual matrices are needed.  (Upper-case Roman indices are vector indices.)
$©$'s and $ú$'s are all just CGW coefficients for $D°D£P\oplus \Upsilon$ and $P°P£\S$ and so carry the indices indicated above.

The anticommutation relations $ÓD,DÕ$
are (where the equation number (\ref{Casey}.D) labels the dimension D):

\begin{align}
Ó D_{à\alpha'} , D_{à\beta'} Õ & = 2P_{à(\alpha'\beta')} \cr
Ó D_{+\alpha'} , D_{-\beta'} Õ & = 2P_{0(\alpha'\beta')}
\end{align}
\begin{align}
Ó D_{\alpha\alpha'} , D_{\beta\beta'} Õ & = 2¶_{ab}P_{(Œº)} \cr
Ó D_{Œ{\bar \alpha}'} , D_{º{\bar \beta}'} Õ & = 2¶_{{\bar \alpha \bar \beta}}\bar P_{(Œº)} \cr
Ó D_{\alpha\alpha'} , D_{º{\bar \beta}'} Õ & = 2C_{Œº}ç_{\alpha'{\bar \beta}'} + 2\di_{(Œº)\alpha'{\bar \beta}'}
\end{align}
\begin{equation}
Ó D_{\alpha\alpha'} , D_{\beta\beta'} Õ = 2¶_{Œ'º'}P_{(Œº)} + 2C_{Œº}ç_{[Œ'º']} + 2\di_{ҌºÔ [Œ'º']}
\end{equation}
\begin{align}
Ó D_{\alpha\alpha'} , D_{\beta\beta'} Õ & = C_{Œº}ç_{[Œ'º']} + \di_{ҌºÔ [Œ'º']} \cr
Ó ÐD{}_{ÀŒ}{}^{Œ'} , ÐD{}_{Àº}{}^{º'} Õ & = ÐC_{ÀŒÀº}Ñç{}^{[Œ'º']} + Ð\di{}_{ÒÀŒÀºÔ}{}^{[Œ'º']} \cr
Ó D_{\alpha\alpha'} , ÐD{}_{Àº}{}^{º'} Õ & = ¶_{Œ'}^{º'} P_{ŒÀº}
\end{align}
\begin{equation}
Ó D_{\alpha\alpha'} , D_{\beta\beta'} Õ = 2C_{Œ'º'}P_{ҌºÔ} + 2C_{Œº}ç_{Ҍ'º'Ô} + 2\di_{ҌºÔ Ҍ'º'Ô}
\end{equation}
\begin{align}
Ó D_{\alpha\alpha'} , D_{\beta\beta'} Õ & = 2C_{Œ'º'}P_{[Œº]} \cr
Ó D^Œ{}_{{\bar Œ}'} , D^º{}_{{\bar º}'} Õ & = 2C_{{\bar Œ}'{\bar º}'}P^{[Œº]} \cr
Ó D_{\alpha\alpha'} , D^º{}_{{\bar º}'} Õ & = 2¶_Œ^º ç_{Œ'{\bar º}'} + 2\di_{Œ}{}^º{}_{Œ'{\bar º}'} .
\end{align}

\noindent (Factors of $¶^{{\rm d}-1}(§)$ on the right-hand side are implicit.  
We use conventions where commutators in a real basis introduce a factor of 2, but not in a complex basis, as in $Ó©,©Õ=2$ but $Óa,aÿÕ=1$.
The symmetrizations aren't intended to introduce extra factors of 2, but only as reminders of the symmetries.  Explicit $©'$ matrices for H$'$=SO(7,8,9) have been omitted.)  
For D=1 we combined $X$ and $(Y,\widetilde Y)$, thus $P$ and $(\Upsilon,\di)$ (so H$'$=SO(9,1)), since in that case they're the same types of fields.  These anticommutators have the interesting feature that they look exactly like those for (simple and extended) super Yang-Mills in the corresponding dimensions 
(at least for D$>$1)
if we take $\Upsilon$ as the scalars and set $\di$=0 except for the fact that the spinor indices $Œ$ have twice the range.

The symmetry is increased for the corresponding superparticle algebra (i.e., ignoring $ú$)
by again combining $\Upsilon$ and $\di$ into $\mho$ (\ref{Larry}).
The result is

\begin{equation}
Ó D_{ŒŒ'} , D_{\beta\beta'} Õ = 2P_{(Œº)(Œ'º')}
\end{equation}
\begin{align}
Ó D_{ŒŒ'} , D_{\beta\beta'} Õ & = 2¶_{Œ'º'}P_{(Œº)} \cr
Ó D_{{\bar Œ}Ќ'} , D_{{\bar º}к'} Õ & = 2¶_{{\bar Œ}'{\bar º}'}P_{(Œ'º')} \cr
Ó D_{ŒŒ'} , D_{{\bar º}{\bar º}'} Õ & = 2\mho_{Œ{\bar º} Œ'{\bar º}'}
\end{align}
\begin{equation}
Ó D_{ŒŒ'} , D_{\beta\beta'} Õ = 2¶_{Œ'º'}P_{(Œº)} + 2\mho_{[Œº] [Œ'º']}
\end{equation}
\begin{align}
Ó D_{ŒŒ'} , D_{\beta\beta'} Õ & = \mho_{[Œº] [Œ'º']} \cr
Ó ÐD_{ÀŒ}{}^{Œ'} , ÐD_{Àº}{}^{º'} Õ & = Ñ\mho{}_{[ÀŒÀº]}{}^{[Œ'º']} \cr
Ó D_{ŒŒ'} , ÐD_{Àº}{}^{º'} Õ & = ¶_{Œ'}^{º'} P_{ŒÀº}
\end{align}
\begin{equation}
Ó D_{ŒŒ'} , D_{\beta\beta'} Õ = 2C_{Œ'º'}P_{ҌºÔ} + 2\mho_{[Œº] Ҍ'º'Ô}
\end{equation}
\begin{align}
Ó D_{ŒŒ'} , D_{\beta\beta'} Õ & = 2C_{Œ'º'}P_{[Œº]} \cr
Ó D^{\bar Œ}{}_{{\bar Œ}'} , D^{\bar º}{}_{{\bar º}'} Õ & = 2C_{{\bar Œ}'{\bar º}'}P^{[\bar Œ\bar º]} \cr
Ó D_{ŒŒ'} , D^{\bar º}{}_{{\bar º}'} Õ & = 2\mho_{Œ}{}^{\bar º}{}_{Œ'{\bar º}'} .
\end{align}

This has allowed us to generalize the group H to the group L for $ÓD,DÕ$ (except for D=5)
as indicated in the above tables.
(When including the dual form $\widetilde Y$, $\Upsilon$ and $\di$ are the currents for $Y$ and $\widetilde Y$, so combining them is natural:  The transverse part of an $n$-form and the transverse part of the dual form combine into a full $(n+1)$-form.)

\section{Fierz}
\label{Fred}


In the previous section we listed all the components of $(©^{\un A})_{{\un \alpha}{\un \beta}}$ for all cases $\mathrm D = 1,\dots,6$.  We now give all the components of $ú_{{\un A}{\un B}{c}}$; between the two, the complete algebra is determined.  The former was given by $ÓD,DÕ$ (superparticle), the latter is given by $[\P,\P]$ (bosonic F-theory):

\begin{equation}
[ P_{0a'} , P_{àb'} ] = à2i¶_{a'b'}»_à
\end{equation}
\begin{align}
[ P_{Œº} , ÐP^{©¶} ] &= 2i¶_{(Œ}^{(©} »_{º)}{}^{¶)} \cr
[ \Upsilon_{a'} , \di_{Œº b'} ] &= 2i¶_{a'b'}»_{Œº}
\end{align}
\begin{align}
[ P_{Œº} , P^{©¶} ] &= 2i¶_{(Œ}^{(©} »_{º)}{}^{¶)} \cr
[ \Upsilon_{a'} , \di_{Œº b'} ] &= 2i¶_{a'b'}»_{Œº}
\end{align}
\begin{align}
[ P_{ŒÀº} , P_{©À¶} ] &= 2i(C_{Œ©}л_{ÀºÀ¶} + ÐC_{ÀºÀ¶}»_{Œ©}) \cr
[ \Upsilon_{Œ'º'} , Ð\di{}_{ÀŒÀº}{}^{©'¶'} ] &= i¶_{[Œ'}^{©'} ¶_{º']}^{¶'} л_{ÀŒÀº} \cr
[ Ñ\Upsilon{}^{Œ'º'} , \di{}_{Œº ©'¶'} ] &= i¶_{[©'}^{Œ'} ¶_{¶']}^{º'} »_{Œº}
\end{align}
\begin{align}
[ P_{Œº} , P^{©¶} ] &= 2i¶_{Ҍ}^{Ò©} »_{ºÔ}{}^{¶Ô} \cr
[ \Upsilon_{Œ'º'} , \di_{Œº}{}^{©'¶'} ] &= 2i¶_{Ҍ'}^{©'} ¶_{º'Ô}^{¶'} »_{Œº}
\end{align}
\begin{align}
[ P_{Œº} , P^{©¶} ] &= 2i¶_{[Œ}^{[©} »_{º]}{}^{¶]} \cr
[ P_{Œº} , P_{©¶} ] &= 2i»_{Œº©¶} \cr
[ P^{Œº} , P^{©¶} ] &= 2i»^{Œº©¶} \cr
[ \Upsilon_{Œ'{\tilde º}'} , \di_Œ{}^º{}_{©'{\tilde ¶}'} ] &= 2iC_{Œ'©'}C_{{\tilde º}'{\tilde ¶}'}»_Œ{}^º \cr
[ \Upsilon_{Œ'{\tilde º}'} , \di_{Œº©¶, ©'{\tilde ¶}'} ] &= 2iC_{Œ'©'}C_{{\tilde º}'{\tilde ¶}'}»_{Œº©¶}  ,
\end{align}
again with equation (\ref{Fred}.D) corresponding to dimension D. 
(We have again omitted the $¶$'s, on which $»$ acts, and used real/complex conventions for factors of 2.  Missing commutators vanish.)

The consistency of the above constructions leaves only the Fierz identity (\ref{Fierz}) to be checked.  (It may be computationally simpler to check the Jacobi (\ref{Jacob}) directly, or the $ÓD,DÕ$ commutator (\ref{DD}) in some representation, especially (\ref{rap}).
The twistor representation (\ref{Chubby}) is also useful.)
This restricts D$_{tot}$=D+D$'$ = 3,4,6,10.  The simplest example is D=1, which has the Fierz identities
\begin{equation}
\delta_{a'b'} \, (©^{\pm a'})_{(àŒ'àº'}(©^{0b'})_{à©')¦¶'} = 0 .
\end{equation}
After making the identifications
\begin{equation}
(©^{0a'})_{àŒ'¦º'} = (©^{\pm a'})_{àŒ'àº'} = (©^{a'})_{Œ'º'} ,
\end{equation}
this gives the standard Fierz for the above dimensions
\begin{equation}
\label{E:Fierz1}
(©_{a'})_{(Œ'º'}(©^{a'})_{©')¶'} = 0 .
\end{equation}
Since all spinor indices are the same, this would require Type IIB.  But if we return to the $X¢Y$ separation, where H$'$=SO(D$_{tot}-$1), there is no IIA/B distinction.

For D=2, one obtains directly the Fierz identity for SO(D$'$) in D$'$=1,2,4,8
\begin{equation}
\label{E:Fierz2}
(©_{a'})_{(Œ'|{\tilde º}'} (©^{a'})_{|©'){\tilde ¶}'} = 2¶_{Œ'©'}¶_{{\tilde º}'{\tilde ¶}'}
\end{equation}
which follows from the previous equation (\ref{E:Fierz1}) upon reduction by 1 space and 1 time dimension:
\begin{align}
a'£(à,a')¼,â
Œ'£(Œ',{\tilde Œ}')¼;â
(©^+)_{Œ'º'}£¶_{Œ'º'}¼,â
(©^-)_{Œ'º'}£¶_{{\tilde Œ}'{\tilde º}'}¼,â \cr
(©^{a'})_{Œ'º'} £ \left( (©^{a'})_{Œ'{\tilde º}'}, (©^{a'})_{º' {\tilde Œ}'}\right) .
\end{align}
(In D$'$=8 it's related to $Ó©,©Õ=2I$ by triality.  The usual identities with spinor indices are used in performing manipulations.  Or one can notice that the equations have no traceless pieces, and therefore take all possible traces that reduce to equations with at least one term with no $¶$ nor $C$.)

For D=3, the direct result is
\begin{equation}
(©_{a'})_{(Œ'|º'}(©^{a'})_{|©')¶'} = 2¶_{Œ'©'}¶_{º'¶'} - ¶_{º'(Œ'}¶_{©')¶'}
\end{equation}
which follows from the D=2 result (\ref{E:Fierz2}) upon reduction by 1 space dimension:
\begin{equation}
a'£(-1,a')¼,â
Œ'£Œ'¼,â
{\tilde Œ}' £Œ'¼;â
(©^{-1})_{Œ'º'}£¶_{Œ'º'}¼,â
(©^{a'})_{Œ'{\tilde º}'} £ (©^{a'})_{[Œ'º']} .
\end{equation}
It's satisfied in D$'$=0,1,3,7.

For D=4, one finds
\begin{equation}
(©_{a'})_{Œ'º'} (Щ{}^{a'})^{©'¶'} = ¶_{[Œ'}^{©'} ¶_{º']}^{¶'} .
\end{equation}
This is satisfied for D$'$=0,2,6, corresponding to the range of $Œ'$ being 1,2,4:  For 0, trivially 0=0.  For 2, trivially 1=1, for 1 complex dimension.  For 6, a vector is an arbitrary antisymmetric tensor, satisfying a reality condition using $·_{abcd}$.
This follows from dimensional reduction as
\begin{align}
a'£(-1,a')¼,â
Œ'£({}_{Œ'},{}^{Œ'})¼;â
(©^{-1})_{Œ'º'}£(¶_{º'}^{Œ'},-¶_{Œ'}^{º'})¼,â
\cr
(©^{a'})_{Œ'º'} £ \left( (©^{a'})_{[Œ'º']}, (Щ{}^{a'})^{[Œ'º']} \right) .
\end{align}

As noted for $ÓD,DÕ$ (\ref{Casey}.D), the Fierz identities appear in the same way as in the various super Yang-Mills theories in D dimensions after reduction to internal spinor indices, in spite of the fact that the starting point (\ref{Fierz}) involves a different ``metric''.  This isn't too surprising considering that F-theory reduces to S-theory, which has the same identities for each chirality.

\section{Prospects}
\label{Conclusions}

The current superalgebra + constraints we have described are sufficient to give a Hamiltonian formulation of F-theory (0-modes and excitations).
A straightforward application is to massless backgrounds.  This would give directly the U-covariant form of the field equations (torsion and curvature constraints) for F-supergravity in F-superspace.  A related problem is to consider such backgrounds for specific solutions, such as AdS$_5ð$S$^5$.

The twistor solution to the section condition suggests a new formulation of F-supergravity where U-duality is manifest even after sectioning.  
Perhaps this approach could be extended to the non-0-modes.
The appearance of pure spinors as gauge parameters for the twistors 
may then imply
a natural explanation for their appearance elsewhere in superstring theory as first-quantized ghosts.

In future papers we plan to elaborate on the cases D=5,6.  For D=7, some conceptual problems must first be solved:  $P$ is expected to form the {\bf 248} representation of E$_8$ (G for that case), while $D$ is in the {\bf 16}$¢{\bf Ñ{16}}$ of SO*(16) (H).  But $ÓD,DÕ$ can give only {\bf 1}, {\bf 120}, and {\bf 135} of SO*(16), while the {\bf 248} decomposes into the adjoint {\bf 120} and spinor {\bf 128}:  There is no way to get a spinor ({\bf 128}) from vectors ({\bf 16}).  (Also, the next largest representation of E$_8$ is the {\bf 3875}, and $ÓD,DÕ$ has only 528 components.)


Another goal is to give the Lagrangian formulation of the supersymmetric theories, as we have for the bosonic ones.
One problem that has not been addressed is the inclusion of the remaining Virasoro operators $\T$, associated with $ $ development.  (We have given only $\S$, corresponding to $»/»§$.)  This is directly related to several other omissions:  the worldvolume metric field, which couples to $\T$ (and how a ``conformal gauge" can be chosen for it), and $û$ symmetry (which closes on $\T$ as well as $\S$).
A better understanding of the $D$ constraint might resolve this issue since $û$ symmetry is a first-class subset of these mixed first and second-class constraints, and because for maximal supersymmetry the massless background for $D$'s current algebra necessarily satisfies the (superspace formulation of F-)supergravity field equations.  

\section*{Acknowledgments}
W{\sc dl}3 is partially supported by the U{\sc mcp} Center for String \& Particle Theory and National Science Foundation grants PHY-0652983 and PHY-0354401. 
W{\sc s} is supported in part by National Science Foundation grant PHY-1316617. 




{\small
\linespread{1}
\selectfont
\bibliography{/Users/wdlinch3/Dropbox/Rashoumon/LaTeX/BibTex/BibTex}

\begin{thebibliography}{10}

\bibitem{Dirac:1962iy}
Paul~A.M. Dirac.
\newblock {An Extensible model of the electron}.
\newblock {\em Proc.Roy.Soc.Lond.}, A268:57--67, 1962.
\newblock \href{http://inspirehep.net/record/8639?ln=en}{[{\sc in}SPIRE
  entry]}.

\bibitem{Polyakov:1986cs}
Alexander~M. Polyakov.
\newblock {Fine Structure of Strings}.
\newblock {\em Nucl.Phys.}, B268:406--412, 1986.
\newblock \href{http://inspirehep.net/record/231722?ln=en}{[{\sc in}SPIRE
  entry]}.

\bibitem{Bergshoeff:1987cm}
E.~Bergshoeff, E.~Sezgin, and P.K. Townsend.
\newblock {Supermembranes and Eleven-Dimensional Supergravity}.
\newblock {\em Phys.Lett.}, B189:75--78, 1987.
\newblock \href{http://inspirehep.net/record/248230?ln=en}{[{\sc in}SPIRE
  entry]}.

\bibitem{deWit:1988ig}
B.~de~Wit, J.~Hoppe, and H.~Nicolai.
\newblock On the quantum mechanics of supermembranes.
\newblock {\em Nucl.Phys.}, B305:545, 1988.
\newblock \href{http://inspirehep.net/record/261702?ln=en}{[{\sc in}SPIRE
  entry]}.

\bibitem{Mezincescu:1987kj}
Luca Mezin\c{c}escu, Rafael~I. Nepomechie, and P.~van Nieuwenhuizen.
\newblock {Do Supermembranes Contain Massless Particles?}
\newblock {\em Nucl.Phys.}, B309:317, 1988.
\newblock \href{http://inspirehep.net/record/22276?ln=en}{[{\sc in}SPIRE
  entry]}.

\bibitem{Brink:1976sc}
L.~Brink, P.~Di~Vecchia, and Paul~S. Howe.
\newblock {A Locally Supersymmetric and Reparametrization Invariant Action for
  the Spinning String}.
\newblock {\em Phys.Lett.}, B65:471--474, 1976.
\newblock \href{http://inspirehep.net/record/109966?ln=en}{[{\sc in}SPIRE
  entry]}.

\bibitem{Deser:1976rb}
Stanley Deser and B.~Zumino.
\newblock {A Complete Action for the Spinning String}.
\newblock {\em Phys.Lett.}, B65:369--373, 1976.
\newblock \href{http://inspirehep.net/record/109946?ln=en}{[{\sc in}SPIRE
  entry]}.

\bibitem{Howe:1977hp}
Paul~S. Howe and R.W. Tucker.
\newblock {A Locally Supersymmetric and Reparametrization Invariant Action for
  a Spinning Membrane}.
\newblock {\em J.Phys.}, A10:L155--L158, 1977.
\newblock \href{http://inspirehep.net/record/5325?ln=en}{[{\sc in}SPIRE
  entry]}.

\bibitem{Lindstrom:1987cv}
U.~Lindstr{\"o}m.
\newblock {First Order Actions for Gravitational Systems, Strings and
  Membranes}.
\newblock {\em Int.J.Mod.Phys.}, A3:2401, 1988.
\newblock \href{http://inspirehep.net/record/22457}{[{\sc in}SPIRE entry]}.

\bibitem{Curtright:1982gt}
Thomas~L. Curtright and Charles~B. Thorn.
\newblock {Conformally Invariant Quantization of the Liouville Theory}.
\newblock {\em Phys.Rev.Lett.}, 48:1309, 1982.
\newblock \href{http://inspirehep.net/record/11757}{[{\sc in}SPIRE entry]}.

\bibitem{Lovelace:1971fa}
C.~Lovelace.
\newblock {Pomeron form-factors and dual Regge cuts}.
\newblock {\em Phys.Lett.}, B34:500--506, 1971.
\newblock \href{http://inspirehep.net/record/67292?ln=en}{[{\sc in}SPIRE
  entry]}.

\bibitem{Linch:2015fya}
William~D Linch and Warren Siegel.
\newblock {F-theory from Fundamental Five-branes}.
\newblock 2015.
\newblock \href{http://arxiv.org/abs/1502.00510}{[arXiv:1502.00510]}.

\bibitem{Linch:2015qva}
William~D. Linch and Warren Siegel.
\newblock {F-theory with Worldvolume Sectioning}.
\newblock 2015.
\newblock \href{http://arxiv.org/abs/1503.00940}{[arXiv:1503.00940]}.

\bibitem{Siegel:1983es}
Warren Siegel.
\newblock {Manifest Lorentz Invariance Sometimes Requires Nonlinearity}.
\newblock {\em Nucl.Phys.}, B238:307, 1984.
\newblock \href{http://inspirehep.net/record/192966}{[{\sc in}SPIRE entry]}.

\bibitem{Cremmer:1979up}
E.~Cremmer and B.~Julia.
\newblock {The SO(8) Supergravity}.
\newblock {\em Nucl.Phys.}, B159:141, 1979.
\newblock \href{http://inspirehep.net/record/140465?ln=en}{[{\sc in}SPIRE
  entry]}.

\bibitem{Cremmer:1978ds}
E.~Cremmer and B.~Julia.
\newblock {The N=8 Supergravity Theory. 1. The Lagrangian}.
\newblock {\em Phys.Lett.}, B80:48, 1978.
\newblock \href{http://inspirehep.net/record/131572?ln=en}{[{\sc in}SPIRE
  entry]}.

\bibitem{Julia:1980gr}
B.~Julia.
\newblock {Group Disintegrations}.
\newblock {\em Conf.Proc.}, C8006162:331--350, 1980.
\newblock \href{http://inspirehep.net/record/155019}{[{\sc in}SPIRE entry]}.

\bibitem{Siegel:1980bp}
W.~Siegel.
\newblock {On-shell O($N$) Supergravity in Superspace}.
\newblock {\em Nucl.Phys.}, B177:325, 1981.
\newblock \href{http://inspirehep.net/record/154313?ln=en}{[{\sc in}SPIRE
  entry]}.

\bibitem{West:2001as}
Peter~C. West.
\newblock {$E_{11}$ and M theory}.
\newblock {\em Class.Quant.Grav.}, 18:4443--4460, 2001.
\newblock \href{http://arxiv.org/abs/hep-th/0104081}{[hep-th/0104081]}.

\bibitem{Hull:2007zu}
C.M. Hull.
\newblock {Generalised Geometry for M-Theory}.
\newblock {\em JHEP}, 0707:079, 2007.
\newblock \href{http://arxiv.org/abs/hep-th/0701203v1}{[hep-th/0701203v1]}.

\bibitem{Berman:2010is}
David~S. Berman and Malcolm~J. Perry.
\newblock {Generalized Geometry and M theory}.
\newblock {\em JHEP}, 1106:074, 2011.
\newblock \href{http://lanl.arxiv.org/abs/1008.1763v4}{[arXiv:1008.1763v4]}.

\bibitem{Coimbra:2011ky}
Andr{\'e} Coimbra, Charles Strickland-Constable, and Daniel Waldram.
\newblock {$E_{d(d)} \times \mathbb{R}^+$ generalised geometry, connections and
  M theory}.
\newblock {\em JHEP}, 1402:054, 2014.
\newblock \href{http://arxiv.org/abs/1112.3989v2}{[arXiv:1112.3989v2]}.

\bibitem{Berman:2012vc}
David~S. Berman, Martin Cederwall, Axel Kleinschmidt, and Daniel~C. Thompson.
\newblock {The gauge structure of generalised diffeomorphisms}.
\newblock {\em JHEP}, 1301:064, 2013.
\newblock \href{http://arxiv.org/abs/1208.5884v2}{[arXiv:1208.5884v2]}.

\bibitem{Siegel:1993xq}
W.~Siegel.
\newblock {Two vierbein formalism for string inspired axionic gravity}.
\newblock {\em Phys.Rev.}, D47:5453--5459, 1993.
\newblock \href{http://arxiv.org/abs/hep-th/9302036}{[hep-th/9302036]}.

\bibitem{Siegel:1993th}
W.~Siegel.
\newblock {Superspace duality in low-energy superstrings}.
\newblock {\em Phys.Rev.}, D48:2826--2837, 1993.
\newblock \href{http://arxiv.org/abs/hep-th/9305073}{[hep-th/9305073]}.

\bibitem{Siegel:1993bj}
W.~Siegel.
\newblock {Manifest duality in low-energy superstrings}.
\newblock {\em Proc. of the Conference Strings '93, Berkeley, CA (World
  Scientific)}, pages 353--363, May 24-29 1993.
\newblock \href{http://arxiv.org/abs/hep-th/9308133}{[hep-th/9308133]}.

\bibitem{Hull:1998br}
C.M. Hull and B.~Julia.
\newblock {Duality and moduli spaces for timelike reductions}.
\newblock {\em Nucl.Phys.}, B534:250--260, 1998.
\newblock \href{http://arxiv.org/abs/hep-th/9803239}{[hep-th/9803239]}.

\bibitem{Linch:2015lwa}
William~D. Linch and Warren Siegel.
\newblock {F-theory Superspace}.
\newblock 2015.
\newblock \href{http://arxiv.org/abs/1501.02761}{[arXiv:1501.02761]}.

\bibitem{Hohm:2013pua}
Olaf Hohm and Henning Samtleben.
\newblock {Exceptional Form of D=11 Supergravity}.
\newblock {\em Phys.Rev.Lett.}, 111:231601, 2013.
\newblock \href{http://arxiv.org/abs/arXiv:1308.1673}{[arXiv:1308.1673]}.

\bibitem{Hohm:2013vpa}
Olaf Hohm and Henning Samtleben.
\newblock {Exceptional Field Theory I: $E_{6(6)}$ covariant Form of M-Theory
  and Type IIB}.
\newblock {\em Phys.Rev.}, D89:066016, 2014.
\newblock \href{http://arxiv.org/abs/arXiv:1312.0614}{[arXiv:1312.0614]}.

\bibitem{Hohm:2013uia}
Olaf Hohm and Henning Samtleben.
\newblock {Exceptional Field Theory II: E$_{7(7)}$}.
\newblock {\em Phys.Rev.}, D89:066017, 2014.
\newblock \href{http://arxiv.org/abs/arXiv:1312.4542}{[arXiv:1312.4542]}.

\bibitem{Godazgar:2014nqa}
Hadi Godazgar, Mahdi Godazgar, Olaf Hohm, Hermann Nicolai, and Henning
  Samtleben.
\newblock {Supersymmetric E$_{7(7)}$ Exceptional Field Theory}.
\newblock {\em JHEP}, 1409:044, 2014.
\newblock \href{http://arxiv.org/abs/arXiv:1406.3235}{[arXiv:1406.3235]}.

\bibitem{Hohm:2014fxa}
Olaf Hohm and Henning Samtleben.
\newblock {Exceptional Field Theory III: E$_{8(8)}$}.
\newblock {\em Phys.Rev.}, D90:066002, 2014.
\newblock \href{http://arxiv.org/abs/arXiv:1406.3348}{[arXiv:1406.3348]}.

\bibitem{Musaev:2014lna}
Edvard Musaev and Henning Samtleben.
\newblock {Fermions and Supersymmetry in $\rm E_{6(6)}$ Exceptional Field
  Theory}.
\newblock {\em JHEP}, 03:027, 2015.
\newblock \href{http://arxiv.org/abs/1412.7286}{[arXiv:1412.7286]}.

\bibitem{West:2003fc}
Peter~C. West.
\newblock {E(11), SL(32) and central charges}.
\newblock {\em Phys.Lett.}, B575:333--342, 2003.
\newblock \href{http://arxiv.org/abs/hep-th/0307098}{[hep-th/0307098]}.

\bibitem{Cremmer:1998px}
E.~Cremmer, B.~Julia, Hong Lu, and C.N. Pope.
\newblock {Dualization of dualities II: Twisted self-duality of doubled fields,
  and superdualities}.
\newblock {\em Nucl.Phys.}, B535:242--292, 1998.
\newblock \href{http://arxiv.org/abs/hep-th/9806106}{[hep-th/9806106]}.

\bibitem{Tseytlin:1990nb}
Arkady~A. Tseytlin.
\newblock {Duality Symmetric Formulation of String World Sheet Dynamics}.
\newblock {\em Phys.Lett.}, B242:163--174, 1990.
\newblock \href{http://inspirehep.net/record/295208?ln=en}{[{\sc in}SPIRE
  entry]}.

\bibitem{Tseytlin:1990va}
Arkady~A. Tseytlin.
\newblock {Duality symmetric closed string theory and interacting chiral
  scalars}.
\newblock {\em Nucl.Phys.}, B350:395--440, 1991.
\newblock \href{http://inspirehep.net/record/297160?ln=en}{[{\sc in}SPIRE
  entry]}.

\bibitem{Siegel:1985ra}
Warren Siegel.
\newblock {Covariant Approach to Superstrings}.
\newblock In W.A. Bardeen and A.R. White, editors, {\em Symposium on anomalies,
  geometry, topology}, page 348, Chicago, March 27-30 1985. World Scientific,
  Singapore.
\newblock \href{http://inspirehep.net/record/213561?ln=en}{[{\sc in}SPIRE
  entry]}.

\bibitem{Siegel:1985xj}
Warren Siegel.
\newblock {Classical Superstring Mechanics}.
\newblock {\em Nucl.Phys.}, B263:93, 1986.
\newblock \href{http://inspirehep.net/record/213958?ln=en}{[{\sc in}SPIRE
  entry]}.

\bibitem{Hatsuda:2014aza}
Machiko Hatsuda, Kiyoshi Kamimura, and Warren Siegel.
\newblock {Ramond-Ramond gauge fields in superspace with manifest T-duality}.
\newblock 2014.
\newblock \href{http://arxiv.org/abs/1411.2206}{[arXiv:1411.2206]}.

\end{thebibliography}
\bibliographystyle{unsrt}
} 

\end{document}